
\hsize=6.5truein
\hoffset=.3truein
\vsize=8.9truein
\voffset=.1truein
\font\twelverm=cmr10 scaled 1200    \font\twelvei=cmmi10 scaled 1200
\font\twelvesy=cmsy10 scaled 1200   \font\twelveex=cmex10 scaled 1200
\font\twelvebf=cmbx10 scaled 1200   \font\twelvesl=cmsl10 scaled 1200
\font\twelvett=cmtt10 scaled 1200   \font\twelveit=cmti10 scaled 1200
\skewchar\twelvei='177   \skewchar\twelvesy='60
\def\twelvepoint{\normalbaselineskip=14pt
  \abovedisplayskip 12.4pt plus 3pt minus 9pt
  \belowdisplayskip 12.4pt plus 3pt minus 9pt
  \abovedisplayshortskip 0pt plus 3pt
  \belowdisplayshortskip 7.2pt plus 3pt minus 4pt
  \smallskipamount=3.6pt plus1.2pt minus1.2pt
  \medskipamount=7.2pt plus2.4pt minus2.4pt
  \bigskipamount=14.4pt plus4.8pt minus4.8pt
  \def\rm{\fam0\twelverm}          \def\it{\fam\itfam\twelveit}%
  \def\sl{\fam\slfam\twelvesl}     \def\bf{\fam\bffam\twelvebf}%
  \def\mit{\fam 1}                 \def\cal{\fam 2}%
  \def\tt{\twelvett}
  \textfont0=\twelverm   \scriptfont0=\tenrm   \scriptscriptfont0=\sevenrm
  \textfont1=\twelvei    \scriptfont1=\teni    \scriptscriptfont1=\seveni
  \textfont2=\twelvesy   \scriptfont2=\tensy   \scriptscriptfont2=\sevensy
  \textfont3=\twelveex   \scriptfont3=\twelveex  \scriptscriptfont3=\twelveex
  \textfont\itfam=\twelveit
  \textfont\slfam=\twelvesl
  \textfont\bffam=\twelvebf \scriptfont\bffam=\tenbf
  \scriptscriptfont\bffam=\sevenbf
  \normalbaselines\rm}

\def\beginlinemode{\endmode
  \begingroup\parskip=0pt \obeylines\def\\{\par}\def\endmode{\par\endgroup}}
\def\beginparmode{\endmode
  \begingroup \def\endmode{\par\endgroup}}
\let\endmode=\par
{\obeylines\gdef\
{}}
\def\singlespace{\baselineskip=\normalbaselineskip}
\def\oneandahalfspace{\baselineskip=\normalbaselineskip
  \multiply\baselineskip by 3 \divide\baselineskip by 2}
\def\doublespace{\baselineskip=\normalbaselineskip \multiply\baselineskip by 2}
\newcount\firstpageno
\firstpageno=2
\footline={\ifnum\pageno<\firstpageno{\hfil}\else{\hfil\twelverm\folio\hfil}\fi}
\let\rawfootnote=\footnote              
\def\footnote#1#2{{\rm\singlespace\parindent=0pt\rawfootnote{#1}{#2}}}
\def\raggedcenter{\leftskip=2em plus 12em \rightskip=\leftskip
  \parindent=0pt \parfillskip=0pt \spaceskip=.3333em \xspaceskip=.5em
  \pretolerance=9999 \tolerance=9999
  \hyphenpenalty=9999 \exhyphenpenalty=9999 }
\parskip=\medskipamount
\twelvepoint            
\overfullrule=0pt       
\def\preprintno#1{
 \rightline{\rm #1}}    
\def\author                     
  {\vskip 3pt plus 0.2fill \beginlinemode
   \singlespace \raggedcenter \twelvesc}
\def\affil                      
  {\vskip 3pt plus 0.1fill \beginlinemode
   \oneandahalfspace \raggedcenter \sl}
\def\abstract                   
  {\vskip 3pt plus 0.3fill \beginparmode
   \doublespace \narrower \noindent ABSTRACT: }
\def\endtitlepage               
  {\endpage                     
   \body}
\def\body                       
  {\beginparmode}               

\def\subhead#1{                 
  \vskip 0.1truein             
  {\raggedcenter #1 \par}
   \nobreak\vskip 0.1truein\nobreak}
\def\refto#1{$|{#1}$}           
\def\references                 
  {\subhead{References}         
   \beginparmode
   \frenchspacing \parindent=0pt \leftskip=1truecm
   \parskip=8pt plus 3pt \everypar{\hangindent=\parindent}}
\gdef\refis#1{\indent\hbox to 0pt{\hss#1.~}}    
\gdef\journal#1, #2, #3, 1#4#5#6{               
    {\sl #1~}{\bf #2}, #3, (1#4#5#6)}           
\def\refstylenp{                
  \gdef\refto##1{ [##1]}                                
  \gdef\refis##1{\indent\hbox to 0pt{\hss##1)~}}        
  \gdef\journal##1, ##2, ##3, ##4 {                     
     {\sl ##1~}{\bf ##2~}(##3) ##4 }}
\def\refstyleprnp{              
  \gdef\refto##1{ [##1]}                                
  \gdef\refis##1{\indent\hbox to 0pt{\hss##1)~}}        
  \gdef\journal##1, ##2, ##3, 1##4##5##6{               
    {\sl ##1~}{\bf ##2~}(1##4##5##6) ##3}}
\def\pr{\journal Phys. Rev., }

\def\prl{\journal Phys. Rev. Lett., }
\def\prpts{\journal Phys. Rep., }
\def\np{\journal Nucl. Phys., }
\def\pl{\journal Phys. Lett., }

\def\endreferences{\body}
\def\endpage                    
  {\vfill\eject}
\def\endpaper                   
  {\endmode\vfill\supereject}
\def\endit
  {\endpaper\end}
\def\ref#1{Ref. #1}                     
\def\Ref#1{Ref. #1}                     

\def\m@th{\mathsurround=0pt }
\font\twelvesc=cmcsc10 scaled 1200
\def\cite#1{{#1}}
\def\(#1){(\call{#1})}
\def\call#1{{#1}}
\def\taghead#1{}
\def\leaderfill{\leaders\hbox to 1em{\hss.\hss}\hfill}
\def\twiddle{\lower.9ex\rlap{$\kern-.1em\scriptstyle\sim$}}
\def\bigtwiddle{\lower1.ex\rlap{$\sim$}}
\def\gtwid{\mathrel{\raise.3ex\hbox{$>$\kern-.75em\lower1ex\hbox{$\sim$}}}}
\def\ltwid{\mathrel{\raise.3ex\hbox{$<$\kern-.75em\lower1ex\hbox{$\sim$}}}}
\def\square{\kern1pt\vbox{\hrule height 1.2pt\hbox{\vrule width 1.2pt\hskip 3pt
   \vbox{\vskip 6pt}\hskip 3pt\vrule width 0.6pt}\hrule height 0.6pt}\kern1pt}
\catcode`@=11
\newcount\tagnumber\tagnumber=0

\immediate\newwrite\eqnfile
\newif\if@qnfile\@qnfilefalse
\def\write@qn#1{}
\def\writenew@qn#1{}
\def\w@rnwrite#1{\write@qn{#1}\message{#1}}
\def\@rrwrite#1{\write@qn{#1}\errmessage{#1}}

\def\taghead#1{\gdef\t@ghead{#1}\global\tagnumber=0}
\def\t@ghead{}

\expandafter\def\csname @qnnum-3\endcsname
  {{\t@ghead\advance\tagnumber by -3\relax\number\tagnumber}}
\expandafter\def\csname @qnnum-2\endcsname
  {{\t@ghead\advance\tagnumber by -2\relax\number\tagnumber}}
\expandafter\def\csname @qnnum-1\endcsname
  {{\t@ghead\advance\tagnumber by -1\relax\number\tagnumber}}
\expandafter\def\csname @qnnum0\endcsname
  {\t@ghead\number\tagnumber}
\expandafter\def\csname @qnnum+1\endcsname
  {{\t@ghead\advance\tagnumber by 1\relax\number\tagnumber}}
\expandafter\def\csname @qnnum+2\endcsname
  {{\t@ghead\advance\tagnumber by 2\relax\number\tagnumber}}
\expandafter\def\csname @qnnum+3\endcsname
  {{\t@ghead\advance\tagnumber by 3\relax\number\tagnumber}}

\def\equationfile{%
  \@qnfiletrue\immediate\openout\eqnfile=\jobname.eqn%
  \def\write@qn##1{\if@qnfile\immediate\write\eqnfile{##1}\fi}
  \def\writenew@qn##1{\if@qnfile\immediate\write\eqnfile
    {\noexpand\tag{##1} = (\t@ghead\number\tagnumber)}\fi}
}

\def\callall#1{\xdef#1##1{#1{\noexpand\call{##1}}}}
\def\call#1{\each@rg\callr@nge{#1}}

\def\each@rg#1#2{{\let\thecsname=#1\expandafter\first@rg#2,\end,}}
\def\first@rg#1,{\thecsname{#1}\apply@rg}
\def\apply@rg#1,{\ifx\end#1\let\next=\relax%
\else,\thecsname{#1}\let\next=\apply@rg\fi\next}

\def\callr@nge#1{\calldor@nge#1-\end-}
\def\callr@ngeat#1\end-{#1}
\def\calldor@nge#1-#2-{\ifx\end#2\@qneatspace#1 %
  \else\calll@@p{#1}{#2}\callr@ngeat\fi}
\def\calll@@p#1#2{\ifnum#1>#2{\@rrwrite{Equation range #1-#2\space is bad.}
\errhelp{If you call a series of equations by the notation M-N, then M and
N must be integers, and N must be greater than or equal to M.}}\else%
 {\count0=#1\count1=#2\advance\count1
by1\relax\expandafter\@qncall\the\count0,%
  \loop\advance\count0 by1\relax%
    \ifnum\count0<\count1,\expandafter\@qncall\the\count0,%
  \repeat}\fi}

\def\@qneatspace#1#2 {\@qncall#1#2,}
\def\@qncall#1,{\ifunc@lled{#1}{\def\next{#1}\ifx\next\empty\else
  \w@rnwrite{Equation number \noexpand\(>>#1<<) has not been defined yet.}
  >>#1<<\fi}\else\csname @qnnum#1\endcsname\fi}

\let\eqnono=\eqno
\def\eqno(#1){\tag#1}
\def\tag#1$${\eqnono(\displayt@g#1 )$$}

\def\aligntag#1\endaligntag
  $${\gdef\tag##1\\{&(##1 )\cr}\eqalignno{#1\\}$$
  \gdef\tag##1$${\eqnono(\displayt@g##1 )$$}}

\def\eqalignno#1{\displ@y \tabskip\centering
  \halign to\displaywidth{\hfil$\displaystyle{##}$\tabskip\z@skip
    &$\displaystyle{{}##}$\hfil\tabskip\centering
    &\llap{$\displayt@gpar##$}\tabskip\z@skip\crcr
    #1\crcr}}

\def\displayt@gpar(#1){(\displayt@g#1 )}

\def\displayt@g#1 {\rm\ifunc@lled{#1}\global\advance\tagnumber by1
        {\def\next{#1}\ifx\next\empty\else\expandafter
        \xdef\csname @qnnum#1\endcsname{\t@ghead\number\tagnumber}\fi}%
  \writenew@qn{#1}\t@ghead\number\tagnumber\else
        {\edef\next{\t@ghead\number\tagnumber}%
        \expandafter\ifx\csname @qnnum#1\endcsname\next\else
        \w@rnwrite{Equation \noexpand\tag{#1} is a duplicate number.}\fi}%
  \csname @qnnum#1\endcsname\fi}

\def\ifunc@lled#1{\expandafter\ifx\csname @qnnum#1\endcsname\relax}

\let\@qnend=\end\gdef\end{\if@qnfile
\immediate\write16{Equation numbers written on []\jobname.EQN.}\fi\@qnend}

\catcode`@=12
\refstyleprnp
\catcode`@=11
\newcount\r@fcount \r@fcount=0
\def\refreset{\global\r@fcount=0}
\newcount\r@fcurr
\immediate\newwrite\reffile
\newif\ifr@ffile\r@ffilefalse
\def\w@rnwrite#1{\ifr@ffile\immediate\write\reffile{#1}\fi\message{#1}}

\def\writer@f#1>>{}
\def\referencefile{
  \r@ffiletrue\immediate\openout\reffile=\jobname.ref%
  \def\writer@f##1>>{\ifr@ffile\immediate\write\reffile%
    {\noexpand\refis{##1} = \csname r@fnum##1\endcsname = %
     \expandafter\expandafter\expandafter\strip@t\expandafter%
     \meaning\csname r@ftext\csname r@fnum##1\endcsname\endcsname}\fi}%
  \def\strip@t##1>>{}}

\def\citeall#1{\xdef#1##1{#1{\noexpand\cite{##1}}}}
\def\cite#1{\each@rg\citer@nge{#1}}	

\def\each@rg#1#2{{\let\thecsname=#1\expandafter\first@rg#2,\end,}}
\def\first@rg#1,{\thecsname{#1}\apply@rg}	
\def\apply@rg#1,{\ifx\end#1\let\next=\relax
\else,\thecsname{#1}\let\next=\apply@rg\fi\next}

\def\citer@nge#1{\citedor@nge#1-\end-}	
\def\citer@ngeat#1\end-{#1}
\def\citedor@nge#1-#2-{\ifx\end#2\r@featspace#1 
  \else\citel@@p{#1}{#2}\citer@ngeat\fi}	
\def\citel@@p#1#2{\ifnum#1>#2{\errmessage{Reference range #1-#2\space is bad.}%
    \errhelp{If you cite a series of references by the notation M-N, then M and
    N must be integers, and N must be greater than or equal to M.}}\else%
 {\count0=#1\count1=#2\advance\count1
by1\relax\expandafter\r@fcite\the\count0,%
  \loop\advance\count0 by1\relax
    \ifnum\count0<\count1,\expandafter\r@fcite\the\count0,%
  \repeat}\fi}

\def\r@featspace#1#2 {\r@fcite#1#2,}	
\def\r@fcite#1,{\ifuncit@d{#1}
    \newr@f{#1}%
    \expandafter\gdef\csname r@ftext\number\r@fcount\endcsname%
                     {\message{Reference #1 to be supplied.}%
                      \writer@f#1>>#1 to be supplied.\par}%
 \fi%
 \csname r@fnum#1\endcsname}
\def\ifuncit@d#1{\expandafter\ifx\csname r@fnum#1\endcsname\relax}%
\def\newr@f#1{\global\advance\r@fcount by1%
    \expandafter\xdef\csname r@fnum#1\endcsname{\number\r@fcount}}

\let\r@fis=\refis			
\def\refis#1#2#3\par{\ifuncit@d{#1}
   \newr@f{#1}%
   \w@rnwrite{Reference #1=\number\r@fcount\space is not cited up to now.}\fi%
  \expandafter\gdef\csname r@ftext\csname r@fnum#1\endcsname\endcsname%
  {\writer@f#1>>#2#3\par}}

\def\ignoreuncited{
   \def\refis##1##2##3\par{\ifuncit@d{##1}%
     \else\expandafter\gdef\csname r@ftext\csname
r@fnum##1\endcsname\endcsname%
     {\writer@f##1>>##2##3\par}\fi}}

\def\r@ferr{\endreferences\errmessage{I was expecting to see
\noexpand\endreferences before now;  I have inserted it here.}}
\let\r@ferences=\references
\def\references{\r@ferences\def\endmode{\r@ferr\par\endgroup}}

\let\endr@ferences=\endreferences
\def\endreferences{\r@fcurr=0
  {\loop\ifnum\r@fcurr<\r@fcount
    \advance\r@fcurr by 1\relax\expandafter\r@fis\expandafter{\number\r@fcurr}%
    \csname r@ftext\number\r@fcurr\endcsname%
  \repeat}\gdef\r@ferr{}\global\r@fcount=0\endr@ferences}

\let\r@fend=\endpaper\gdef\endpaper{\ifr@ffile
\immediate\write16{Cross References written on []\jobname.REF.}\fi\r@fend}

\catcode`@=12

\citeall\refto		
\citeall\ref		%
\citeall\Ref		%

\referencefile

\def\oneandthreefifthsspace{\baselineskip=\normalbaselineskip
  \multiply\baselineskip by 8 \divide\baselineskip by 5}

\font\titlefont=cmr10 scaled\magstep3
\def\bigtitle                      
  {\null\vskip 3pt plus 0.2fill
   \beginlinemode \doublespace \raggedcenter \titlefont}

\oneandahalfspace
\body
\preprintno{UFIFT-HEP-94-22}

\bigtitle{\bf SUPERSTRINGS:THE VIEW FROM BELOW}

\author{P.Ramond}
\affil{\sl Institute for Fundamental Theory,
 University of Florida, Gainesville, FL 32611}
\centerline{Invited Lecture at the first G\"ursey Symposium,
Istamboul, June 1994}
\body
\abstract
We review  the Standard Model in a form conducive to formulating its
possible short distance extensions. This depends on the value of the
Higgs mass, the only unknown parameter of
the model. We suggest methods to reproduce  many of the small numbers in
the model in terms of scale ratios, applying see-saw like ideas to the
breaking of chiral symmetries. We then investigate how the $N=1$
Standard Model extrapolated to or near the Planck scale can fit
superstring models, emphasizing the use of some non-renormalizable
operators generic to superstrings.
\endtitlepage
\oneandthreefifthsspace
\subhead{\bf I Introduction}

\noindent The Standard Model is a remarkably compact description of
all fundamental matter,
described in terms of only three gauge groups and nineteen parameters.
Yet, it hardly looks like a fundamental theory; it  probably is the low
energy manifestation of a more integrated, more satisfying, and less
broken-up theory. If it is indeed to be viewed as an effective low
energy theory, there must exist an ultraviolet cut-off. The {\it raison
d'\^ etre} and the value of this cut-off are the central question of
fundamental theory. Since there is no experimental indication of its
existence, it must be at least of the order
of hundreds of GeVs. At the higher end, Nature provides us with its own
cut-off, the Planck scale, the largest cut-off we can presently imagine
to the standard model. However it is far removed from present
experimental scales.

Local field theories of gravity certainly break down
at the Planck scale, and the only known possible cure is to formulate
theories which
deviate from space-time locality, superstring theories. Of
these, the heterotic string seems to contain all the necessary
ingredients needed to reproduce the low energy world. Unfortunately,
there is no detailed matching between the standard model and string
theory. Yet if our world has its origin in a superstring theory, there
ought to be satisfactory matching of the two at some scale below the
Planck scale.
The knowledge of the standard model alone, may not be sufficient
to identify this  match. Still, effective theories
derived from superstrings typically reproduce not only renormalizable
interactions, such as those found in the standard model, but also a
pattern of non-renormalizable interactions, some of which may provide
low energy signatures for superstring models. In addition, these
theories contain not only the observed chiral families, but also a
number of vector-like particles, some with electroweak quantum numbers,
but with hitherto undetermined $\Delta I_W=0$ masses.
Supersymmetry at low energy provides  another hint, although superstring
theories do not yet predict its breaking scale.

We start by reviewing the standard model, and present arguments
for low energy supersymmetry. The $N=1$ standard model is perturbative
until the Planck scale, where, we can hope to match it with
superstring models. We  present a mode of attack which, although
based mostly on
our incomplete knowledge of the Yukawa sector,  hints at the
presence of certain types of non-renormalizable terms, some of which are
generic to superstrings.

\subhead{\bf II The $N=0$ Standard Model}

\noindent Extension of the standard model predicts new phenomena at shorter
distances, although  none so far have distinguished
themselves either by reproducing the {\it values} of the parameters,
or even their multiplicity. Thus it is time to review the types of
extensions which might generically explain the observed patterns.

The $N=0$ standard model is described by three Yang-Mills groups, each
with its own dimensionless gauge coupling, $\alpha_1$ for the
hypercharge $U(1)$, $\alpha_2$ for the   weak isospin $SU(2)$, and
$\alpha_3$ for  QCD. QCD itself predicts strong CP violation, with
strength proportional to a fourth dimensionless parameter
$\overline\theta$.

The electroweak symmetry breaking Higgs sector contains two unknowns, a
dimensionless Higgs self-coupling, and the Higgs mass. The $``$measured"
value of  the Fermi coupling accounts for one parameter, and the other
is the  value of the Higgs mass, the only parameter of the model yet to
be determined from experiment.
The Yukawa interactions between the fermions and the Higgs yields the
nine masses of the elementary fermions. This sector also contains
three mixing angles which monitor
interfamily decays, and one phase which describes CP violation.

It also  contains two dimensionful parameters,
the Higgs mass, and the QCD confinement scale, obtained by
dimensional transmutation.
The QCD scale is a tiny number in Planck units
${\Lambda_{QCD}}\sim 10^{-20} M_{Pl}\ .$
This small number has a natural  explanation
due to  the logarithmic variation  of the QCD coupling with scale.

The Higgs mass is unknown, but  the
electroweak order parameter is determined by the
Fermi constant. In terms of the Planck mass it is also  very small
${G_F^{-1/2}}\sim 10^{-17} M_{Pl}^{}\ .$
The origin of this small number is a matter of much speculation.
In perturbation theory the Higgs mass is of the same order of magnitude
as the electroweak order parameter. The most natural idea is to relate
this number to dimensional transmutation associated with new strong  forces
just beyond electroweak scales, called technicolor.  It yields a
satisfying natural explanation of this value, but these  models fail to
reproduce the values of the fermion masses.

Another class of theories postulates supersymmetry\refto{reviews} at TeV
scales. There, the electroweak order parameter is related to that  of
supersymmetry breaking. While  not at first sight  very
economical, the breaking of  supersymmetry
automatically generates electroweak breaking\refto{trigger} in a wide
class of theories. The beautiful ideas of technicolor can then be applied to
supersymmetry breaking, without encountering the problem of fermion
masses of technicolor applied to electroweak breaking.

There are many other numbers to explain, notably in the Yukawa sector of
the theory. Quark and charged lepton masses break weak isospin by half a unit,
along $\Delta I_W={1\over 2}$, and hypercharge by one unit,  the same
quantum numbers as  the electroweak order parameter,
which also gives the W-boson its mass. In this sense charged fermion
masses should be of the same order as the W mass. This happens only
for the top  quark. The others are unnaturally small
$${m_{u,d}\over M_W}\sim {\cal O}(10^{-4})\ ;\qquad
{m_s\over M_W}\sim {\cal O}(10^{-3})\ ;\qquad {m_c\over M_W}
\sim{\cal O}(10^{-2})\ ;
\qquad {m_b\over M_W}\sim .05\ .$$
Similarly for the charged leptons,
$${m_e\over M_W}\sim {\cal O}(10^{-5})\ ;\qquad
{m_\mu\over M_W}\sim {\cal O}(10^{-3})\ ;\qquad {m_\tau
\over M_W}\sim .02\ ,$$
which  range from the tiny to the small.
Neutrino masses are predicted to be exactly zero in the standard model
only because of the global chiral lepton number symmetries. However there is
mounting experimental evidence that neutrinos have masses. In the
absence of new degrees of freedom they are of the Majorana kind, and
break weak isospin by one unit, as  $\Delta I_W=1$.
Direct experimental limits on neutrino masses indicate that they are at most
extremely small:
${m_{\nu_e}}< 10^{-17} M_W\ .$

The values of the three gauge parameters are known to great accuracy
from measurements at low energy, although because of endemic problems
associated with strong QCD, the color  coupling is the least well known.
Given these parameters, we can extrapolate the standard model to
shorter distances, using the renormalization group perturbatively.
The most interesting
effect occurs in the extrapolation of the three gauge couplings.
The hypercharge and weak isospin couplings meet at a scale
of $10^{13}$ GeV, with a value $\alpha^{-1}\approx 43$, but at that
scale, the QCD coupling is much larger, $\alpha_3^{-1}\approx 38$. Thus,
although the quantum numbers indicate  possible unification into a
larger non-Abelian group, the gauge coupling do not follow suit in this
naive extrapolation. Historically, before the couplings were
measured to such accuracy, it was believed that all three did indeed
unify in the ultraviolet. In the ultraviolet, the values of these
couplings is less
disparate than at experimental scales. Similarly, nothing spectacular
occurs to the Yukawa couplings. For instance, the botton quark and
$\tau$ lepton Yukawa couplings meet around $10^9$ GeV, and part  in
the deeper ultraviolet.

The situation is potentially more extreme in the Higgs sector
because of the renormalization group behavior of the Higgs self
coupling\refto{sher}. We can consider two cases, depending on the value
of the Higgs mass.
If it is below $150$ GeV, the self-coupling turns negative
somewhere below Planck scale. This results in a loss of perturbation
theory, with  a potential  unbounded from below. Using
the recently announced value of the top quark mass,  a Higgs
mass of $120$ GeV  means that  $``$instability" sets in at 1 TeV,
indicating some new physics  at that scale. When operative,
this bound provides a low (with respect to Planck mass) energy cut-off
for the standard model.

If the Higgs mass is above $200$ GeV, its self-coupling rises
dramatically towards its Landau pole at a relatively low energy scale.
It means loss of  perturbative control of the theory, and
sets an upper bound on the Higgs mass since there is no evidence of
any strong electroweak coupling at experimental scales.
 Strong coupling must happen,
meaning that the Higgs is a composite. An example of this
view is the technicolor scenario where the Higgs is a condensate of
techniquarks.

Within  a tiny range of intermediate values for the Higgs mass, the
instability and triviality bounds are pushed to scales beyond the Planck
length, and  there is no standard model prediction of new physics;
the cut-off is indeed the Planck scale.

The dependence of the various standard model parameters on the cut-off is
very different.
Quantum fluctuations {\it additively}
renormalize the Higgs mass with a term linearly proportional to the
cut-off.
Thus even if the Higgs mass is in a region that does not {\it
technically} require new physics below Planck mass, its value is
unnaturally small, if Planck mass is  the cut-off.
On the other hand, the cut-off
dependence of any chiral fermion mass is only
logarithmic. The reason is chiral symmetry, which is recovered by
setting the fermion mass to zero. It affords a protection mechanism
which results in this weak cut-off dependence.

\subhead{\bf III The $N=1$ Standard Model}

\noindent Supersymmetry avoids the {\it technical}
naturalness problem by
linking any fermion to a boson of the same mass. With exact
supersymmetry, the boson mass is then  protected by the chiral symmetry
of the fermion. As long as supersymmetry is broken at energies in the
range of TeV, this is enough protection to produce a low Higgs mass.
This might seem to be small progress, since a new symmetry has been
introduced to relax the strong cut-off dependence, a  symmetry which has to
be broken itself at a small scale, ${V^{}_{SUSY}}\sim 10^{-15} M_{Pl}\ .$

In the $N=1$ standard model, there are only gauge and Yukawa coupling
constants.  None of these couplings blow up below Planck mass. In
particular, the perky Higgs self-coupling is replaced by the square of
gauge and Yukawa couplings, which  allows for the perturbative
extrapolation all the way to Planck scale, opening the way for
comparison with fundamental theory!

There are tantalizing hints of simplicity in the extrapolation of the
couplings. Firstly the gauge couplings seem to be much
closer to unification, and at a scale large enough not to be  invalidated
by proton decay
bounds. The hypercharge and weak isospin couplings meet
at a scale of the order of $10^{16}$ GeV, with a value
$\alpha^{-1}\approx 25$, and the QCD coupling is much
closer to, if not right on the same value\refto{unification}. It may
still be a shade higher than the others, with
$(\alpha^{-1}-\alpha_3^{-1})\le 1.5$.

The second remarkable thing is that with simple boundary conditions at
or near Planck mass, inspired by a simple picture of supersymmetry
breaking, the renormalization group drives one of the Higgs masses to
imaginary values in the infrared. This in turns triggers electroweak
breaking[\cite{trigger}], made possible by the large top quark mass!

The Higgs mass is not arbitrarily high in the minimal
extension. At tree-level, it is predicted to be below the Z-mass, but it
suffers large radiative corrections due to the top Yukawa coupling,
raising it above the Z, but not by an arbitrarily large amount\refto{gordy}.

This general scheme allows us to study the pattern of fermion masses at
these shorter distances; there are more
regularities with supersymmetry. For instance, the bottom
quark and $\tau$ masses seem to unify at or around $10^{16-17}$
GeV\refto{btau}, the same scale where the gauge couplings converge.

The most striking aspect of the fermion masses is that only one
chiral family has large masses, leading us  to consider theories
where the tree-level Yukawa matrices are simply of the form
$${\bf Y}_{u,d,e}=y^{}_{t,b,\tau}\pmatrix{0&0&0\cr 0&0&0\cr
0&0&1\cr}\ .$$
These matrices imply a global chiral symmetry, $U(2)_L\times
U(2)_R$, in each charged sector.
The hierarchy between
the bottom and top quark masses  requires  explanation. In the
$N=1$ model, it is linked  to another parameter which comes from the
Higgs sector, the ratio of the {\it vev} of the two Higgs. Hence it may not
pertain to properties of the Yukawa matrices.
Why are the other two families so light? Starting from the rank two Yukawa
matrices, we must find a scheme by which the zeros get filled, presumably
in higher orders of perturbation theory. In order to see how this might
come about, let us examine one well-known case in
which small numbers are naturally generated, the see-saw
mechanism\refto{gmrsy}.

In the standard model, the neutrino Majorana mass matrix is zero at
tree-level. A detailed examination shows that these zeros are protected
from quantum corrections by conservation of  chiral global lepton number
for each species.
In the see-saw mechanism,  the usual neutrinos are
mixed with new electroweak singlet fields (neutral leptons),
by $\Delta I_w=1/2$ terms, of electroweak breaking strength,
 to give them the same
lepton numbers. These new particles can acquire $\Delta I_W=0$
Majorana masses, $M$, of any magnitude, in particular
well above the electroweak scale.
Upon diagonalization, this generates a mass for
the familiar neutrinos, depressed from typical  electroweak
values by the ratio of scale ${m\over M}$,
where $m$ is the   electroweak order parameter.
A scale ratio between
electroweak and  chiral lepton
number  breaking is used to generate a small number.

A similar analysis may apply to the charged Yukawa matrices, where the
zeros are also protected by chiral symmetries. We first
couple the massless fermions of the first two families to
 new fermions with similar
quantum numbers, thereby extending  the chiral symmetries to them.
Unlike the neutral case, these new fermions have electroweak charges, and
cannot have Majorana masses. Breaking of chiral symmetry is done by
their Dirac masses, which requires the presence of vector-like partners (this
differs from the neutral sector), along the  $\Delta
I_W=0$ direction at a new undetermined scale M.

One may also take the point of view that these non-renormalizable operators
come from physics beyond the Planck scale, in which case, the question is
relegated to one of classifying the possible non-renormalizable
operators.

Now that low energy supersymmetry allows its perturbative
extrapolation deep in the ultraviolet, we may ask how it can be made to
match with more fundamental theories, one type being superstrings.

\subhead{\bf IV. Matching to Superstrings}

\noindent Superstring theories are not understood in detail, but some of
their generic features are evident. We are interested in the
effective theory they generate at or near Planck scale. An estimate of
string effects indicates that this scale is related to the gauge
coupling through the formula [\refto{kounnas}]
$$M_U\approx
2.5\sqrt{\alpha^{}_U}\times 10^{18}\> {\rm GeV}\ .$$
With $M_X=10^{16}$ GeV, and
$\alpha_U^{-1} < \alpha^{-1}_X \approx 25$,
this implies that contact with the superstring can be made provided that
${M_U/M_X}>50$: there is a discrepancy in the matching of scales.

The second feature of superstring theories is to produce at lower energies
 remnants of
$\bf 27$ and $\overline{\bf 27}$ representations of $E_6$. In the effective
low energy theory, these yield  the three chiral families, together
with  many vector-like
particles, capable of sharing quantum numbers with the chiral  particles.
These particles may be used in see-saw like mechanisms to generate small
numbers in the Yukawa matrices.  It also means that there may be many
intermediate thresholds between the supersymmetry and unification scales

A third feature of superstring theories is that the gauged group at the
string scale is larger than the standard model group. This implies the
existence of more gauge bosons at intermediate scales and
many vector-like particles with electroweak quantum numbers. The values of
their masses to be determined by the flat directions in the superpotential,
and the discrete symmetries of the particular model.

A fourth  generic feature is the existence of a local $U(1)$ symmetry,
with anomaly cancelled through the Green-Schwarz mechanism. This symmetry
is however broken close to the Planck scale. Does any trace of this
symmetry appear in the extrapolated low energy standard model? Ib\`a\~nez
[\cite{Ib}]has argued that this symmetry can be used to fix the Weinberg angle
in superstring theories.  Following   Ib\`a\~nez and
Ross[\cite{IR}], we argue[\cite{US}]
that this Abelian symmetry sets the dimensions  of the  Froggart
and Nielsen[\cite{FN}] Yukawa operators.
Are any of these features present in the extrapolated low energy theory?

Consider first the unification of the gauge couplings.
It is predicated on two assumptions: that the
weak hypercharge coupling is normalized to its unification into a higher rank
Lie group, such as $SU(5)$, $SO(10)$ or $E_6$\refto{gut}), and on the absence
of  intermediate thresholds with matter carrying strong or
electroweak quantum numbers between $1$ TeV and $10^{16}$ GeV.
The gauge couplings may not exactly unify at $M_X$, and
we may want to alter this simple picture by requiring
 at least one intermediate threshold between the SUSY
scale and the illusory unification scale at $M_X$ to obtain unification at
the string scale $M_U$[\refto{MR}].

At one-loop, the couplings $
\alpha_i^{-1}(t)$ for the three gauge groups, ($i=1,2,3$ for
$U(1)_Y$, $SU(2)_L$, and $SU(3)^c$, respectively)
run with scale according to
$$=\alpha_i^{-1}(t_X)+{b_i\over 2\pi}(t-t_X)\ ,$$
where
$$
t=\ln (\mu/\mu_0)\ ,\qquad t_X=\ln (M_X/\mu_0)\ ,
$$
and $\mu_0$ is an arbitrary reference energy.
For the three families and two Higgs doublets of  the
minimal supersymmetric standard model, we have
$$
b_1=-{33\over 5}\ ;\qquad b_2=-1\ ;\qquad b_3=3\ .
$$
Since the low energies values of  $\alpha_1$ and $\alpha_2$
are known with the greatest accuracy, we use their trajectories to
define $t_X$ as the scale at which they meet:
$$
\alpha^{-1}_X\equiv\alpha_1^{-1} (t_X) = \alpha_2^{-1} (t_X) \ .
$$
The extrapolated data show that
$\alpha^{-1}_X\approx 25$, with $M_X \approx 10^{16}$ GeV.
We do not assume precisely the
same value for $\alpha_3(t_X)$ at that scale; rather we set
$$
\alpha^{-1}_X = \alpha_3^{-1} (t_X) + \Delta \ .
$$
Present
uncertainties  on the QCD coupling suggest that $$|\Delta | \le 1.5\ .
$$
Suppose there is an intermediate threshold above
supersymmetry at
$$
t_I=\ln({M_I/\mu_0})\ ;\qquad t_I<t_X\ ,
$$
caused by new vector-like
particles with electroweak singlet masses at $M_I$.
Their effect is to alter the $b_i$ coefficients:
$$
b_i\to b_i-\delta_i\ ,~~~~i=1,2,3\ .$$
By requiring unification at $M_U$, we find the constraints
$$
{r\over 14} = {t_U - t_X \over t_U - t_I}
$$
and
$$
{q\over 4} = {t_U - t_X - \pi \Delta/2 \over t_U - t_I}
\>\ ,
$$
written in  terms of
$$q\equiv \delta_3-\delta_2
\qquad {\rm and }\qquad {2\over 5}r\equiv\delta_2-\delta_1\ .$$
For vector-like matter generated from superstrings, $q$ and $r$ are integers.
The value of the gauge coupling at unification is now
$$
\alpha^{-1}_U=\alpha^{-1}_X-{1\over 2 \pi}\left [
\delta_2 (t_U - t_I)+ t_U - t_X \right ]\ .$$
These equations have solutions for  non-exotic matter.
For instance when $\Delta =0.82$ with $r=5$, $q=1$, we get
$$M_U=7.5\times 10^{17} {\rm GeV}\ ;\qquad M_I=4.4\times 10^{12} {\rm
GeV}\ ;\qquad \alpha^{-1}_U=11\ .$$
However most  solutions do not allow large $M_X/M_I$.

In realistic superstring models,
the  assumption of one intermediate scale is probably not justified.
For several intermediate
thresholds, by applying these equations repeatedly,
we obtain similar equations, with $q$ and $r$ replaced by average quantities
which are no longer integers.
Take for instance the interesting example of the 3-family
Calabi-Yau superstring model of ref.~[\cite{GKMR}].
After flux breaking, the surviving gauge group is
$$
SU(3)_L\times SU(3)^c\times SU(3)_R\ .
$$
There are at least two {\it a priori} distinct
intermediate scale order parameters associated with each reduction in rank.
Many chiral superfields  survive flux breaking:
9  leptons,
6  mirror leptons, 7 quarks,
4 mirror quarks, 7 antiquarks , 4 mirror antiquarks.
With  all these particles  concentrated at
one mass scale, there is no solution, but this is hardly realistic.
It is convenient  to define the effective intermediate scale as
the average intermediate scale weighted by $\delta_2$, i.e.,
$$
t_{\overline I} \equiv {\sum_{a=1}^N t_{Ia} \delta_{2a}
\over \sum_{a=1}^N \delta_{2a}} =
{1\over 29}\sum_{a=1}^N t_{Ia} \delta_{2a}
\> .
$$
Taking $5 \times 10^{17}$ GeV as a minimum for $M_U$, we find the high
value.
$$
M_{\overline I} > 3 \times 10^{15} \> {\rm GeV}
$$
Gauge coupling unification can be attained in this example
in a calculably perturbative way, but it requires that many of the
vector-like particle which survive flux breaking
be very close to the string scale, and that electroweak-doublet vectorlike
particles be heavier on average than the strongly-interacting electroweak
singlet vectorlike particles.

It might seem rather surprising
that in the MSSM the gauge couplings
should appear to be nicely headed for unification at $M_X$, only to be
redirected to a new meeting place at $M_U$, but  the apparent perverseness
of this situation allows us put some non-trivial constraints on the
scenario.

Let us now turn to the last topic, the possibility of an
Abelian gauge symmetry, with anomaly cancelled by the Green-Schwarz
mechanism. Such can be recognized if it plays a role  in determining the
dimensions of the entries of the Yukawa matrices[\cite{IR,US}].
The  most general Abelian charge that can be assigned to
the particles of the Minimal
Supersymmetric Standard Model,  with $\mu$ term, can be written as
$$X=X_0+X_3+{\sqrt 3}X_8\ ,$$
where $X_0$ is the family independent part, $X_3$ is along $\lambda_3$,
and $X_8$ is along $\lambda_8$.
We set
$$X_{3,8}^{}=(a^{}_{3,8},b^{}_{3,8},c_{3,8}^{},d_{3,8}^{},e_{3,8}^{})\ ,$$
where the entries correspond to the components in the family space of
the fields ${\bf Q}$, $\overline{\bf u}$, $\overline{\bf d}$, $L$, and
$\overline e$, respectively. Both Higgs doublets have the same zero
X-charge, without loss
of generality, since an imbalance can be created by mixing in the
hypercharge $Y$.

With the tree-level Yukawa coupling {\it only}
to the third family, we obtain the constraints
$${m+n\over 3}=2(a_8+b_8)\ ,\ \ {m+p\over 3}=2(a_8+c_8)\ ,\ \ {q+r\over
3}=2(d_8+e_8)\ .$$
The excess X-charge at each
of their entries is made up by powers of an electroweak singlet field,
resulting in operators of higher dimensions. A typical term would be of
the form
$${\bf Q}_i\overline{\bf u}_jH_u\left({\theta\over M}\right)^{n_{ij}}\ ,$$
where $\theta$ is some field with charge $x$, $M$ is some
large scale, and $n_{ij}$ is the excess X-charge listed above for the
Yukawa matrices.
The exponents are determined by X-symmetry. In order to
produce a small coefficient, the $i$th and $j$th fermions need to go
through a number of intermediate steps to interact. The larger the
number steps, the larger $n_{ij}$, and the smaller the entry in the
effective Yukawa matrix. This approach was advocated long ago by
Froggart and Nielsen\refto{FN}.
This yields  approximate zeros in the matrices,
creating textures[\cite{RRR}] .
For example, in the charge 2/3 sector,
$$n_{12}x=3(a_8+b_8)+a_3-b_3\ .$$
Since $\theta$ may have a large expectation value, it is likely
accompanied by its vector-like partner $\overline\theta$, with opposite
charge, showing that the exponents $n_{ij}$ need not be positive, but
if  all the $n_{ij}$ are positive,
several interesting  phenomenological consequences follow[\refto{US}].
First the $n_{ij}$ exponents are
not all independent,
resulting in order of magnitude estimates among the Yukawa matrix
elements
$$\eqalign{(Y)_{11}&\sim {(Y)_{13}(Y)_{31}\over (Y)_{33}}\ ,\cr
(Y)_{22}&\sim {(Y)_{23}(Y)_{32}\over (Y)_{33}}\ ,\cr   }
$$
valid for each of the three charge sectors. These relations are
consistent with many of the allowed textures. Another
important consequence is that the X-charge of the determinant in each charge
sector is {\it independent} of the texture coefficients that distinguish
between the two lightest families
$${\rm charge}~{2\over 3}~:~6(a_8+b_8)\equiv U\ ,\
{\rm charge}~-{1\over 3}~:~6(a_8+c_8)\equiv D\ ,\
{\rm charge}~-1~:~6(d_8+e_8)\equiv L\ .$$

Let the value of $\theta\over M$ be a small parameter
$\lambda$. In the simplest case, this parameter would be the same for
all three charge sectors. Then we have
$${ m_dm_sm_b\over m_em_\mu m_\tau}\sim {\cal O}(\lambda^{(D-L)/x})\ .$$
It is more difficult to compare the up and down sectors in this way
since we do not know the value of $\tan\beta$, which sets the
normalization between the two sectors
$${ m_um_cm_t\over m_dm_sm_b}\sim \tan^3\beta\times{\cal
O}(\lambda^{(U-D)/x})\ .$$
Since this ratio is much larger
than one, it means either that $\tan\beta$ is itself large, with $U$
close to $D$, or that $\tan\beta$ is not large, but $D>U$.

In general, the X symmetry is anomalous. The three chiral families
contribute to the mixed gauge anomalies as follows
$$\eqalign{C_3&=2m+n+p\ ,\cr
C_2&=3m+q+2s\ ,\cr
C_1&={1\over 3}m+{8\over 3} n+{2\over 3}
p+q+2r+2s\ .\cr}$$
The subscript denotes the gauge group of the Standard Model, {\it i.e.}
$1\sim U(1)$, $2\sim SU(2)$, and $3\sim SU(3)$.
The X-charge also has a mixed gravitational anomaly, which is simply
the trace of the X-charge,
$$C_g=(6m+3n+3p+2q+r+4s)+C_g^\prime\ ,$$
where $C_g^\prime$ is the contribution from the particles that do not
appear in the minimal $N=1$ model.
The last anomaly coefficient is that of the X-charge itself, $C_X$,
which is the sum of the cubes of the X-charge.

It was suggested by Ib\` a\~ nez[\cite{Ib}], that an anomalous $U(1)$
symmetry, with its anomalies cancelled through the Green-Schwarz
mechanism, is capable of relating the ratio of gauge couplings to the
ratios of anomaly coefficients
$${C_i\over k_i}={C_X\over k_X}={C_g\over k_g}\ ,$$
 which relates the Weinberg
angle to the anomaly coefficients, without the use of Grand Unification.
The $k_i$ are the Kac-Moody levels;  are integers for
the non-Abelian factors only. The mixed $YXX$ anomaly, however,
must vanish by itself.

We demand that the non-Abelian gauge groups have the same
Kac-Moody levels, which means that
$$C_2=C_3\qquad {\rm or}\qquad  q=n+p-m-2s\ .$$
Secondly we require that at or near the unification or string scale,
the Weinberg angle have the value
$$\sin^2\theta_W={3\over 8} \ ,$$
which translates into the further constraint
$$5C_2=3C_1\qquad {\rm or}\qquad r=2m-n\ .$$
These equations are sufficient to infer that $L=D$, which implies,
remarkably enough, that the products of the charged lepton masses is of
the same order of magnitude as that of the down-type quarks[\cite{US}].
It is
satisfying to note that extrapolation of the masses to short distances
indicates that these two products are in fact roughly equal in the deep
ultraviolet.

This formalism has been used[\cite{IR}] to generate symmetric textures,
of the kind
found to be allowed by experiment[\cite{RRR}]. Work is in progress to
determine how these equations constrain possible
textures.  One result is that it appear to be difficult to generate
acceptable constraints, without invoking Green-Schwarz cancellation.
In that case, this particular way of generating textures would require
the type of mechanism that is generic to superstrings!

The following examples have shown how several
problems with the standard model might require a
superstring explanation. While it is clearly too soon to claim to have
made the connection, it is a fruitful path to take, as  we must
try to match the apparent unruliness and ugliness of the world we observe
to the more beautiful and satisfying constructs of our imagination. Feza
would have liked that.

I wish to thank Professor Serdaro\u glu for her kind hospitality during
the conference.
This work was supported in part by the United States Department of
Energy under Contract No. DEFG05-86-ER-40272.

\references

\refis{FN} C.~Froggatt and H.~B.~Nielsen \np B147, 277, 1979.

\refis{reviews}
For reviews, see H.~P.~Nilles,  \prpts 110, 1, 1984 and
H.~E.~Haber and G.~L.~Kane, \prpts 117, 75, 1985.

\refis{trigger}
L.~E.~Ib\'a\~nez and  G.~G.~Ross,
\journal Phys.~Lett., 110B, 215, 1982;
K.~Inoue, A.~Kakuto, H.~Komatsu, and S.~Takeshita,
\journal Prog. Theor. Phys., 68, 927, 1982;
L.~Alvarez-Gaum\'e, M.~Claudson, and M.~Wise,
\np B207, 16, 1982;
J.~Ellis, J.~S.~Hagelin, D.~V.~Nanopoulos, and
K.~Tamvakis,
\journal Phys.~Lett., 125B, 275, 1983.

\refis{gut}
J.~C.~Pati and A.~Salam,
\pr D10, 275, 1974;
H.~Georgi and S.~Glashow,
\prl 32, 438, 1974;
H.~Georgi, in {\it Particles and Fields-1974}, edited by C.E.Carlson,
AIP Conference Proceedings No.~23 (American Institute of Physics,
New York, 1975) p.575;
H.~Fritzsch and P.~Minkowski,
\journal Ann.~Phys.~NY, 93, 193, 1975;
F.~G\" ursey, P.~Ramond, and P.~Sikivie,
\pl 60B, 177, 1975.

\refis{btau} H.~Arason, D.~J.~Casta\~no, B.~Keszthelyi, S.~Mikaelian,
E.~J.~Piard, P.~Ramond, and B.~D.~Wright,
\prl 67, 2933, 1991;
A.~Giveon, L.~J.~Hall, and U.~Sarid,
\pl 271B, 138, 1991.

\refis{gordy} G.~L.~Kane, C.~Kolda, and J.~D.~Wells,
\prl 70, 2686, 1993.

\refis{sher}M.~Sher, \prpts 179, 273, 1989.

\refis{RRR}P. Ramond, R.G. Roberts and G. G. Ross, \np B406, 19, 1993.

\refis{Ib}L. Ib\'a\~nez, \pl B303, 55, 1993.

\refis{IR}L. Ib\'a\~nez and G. G. Ross, \pl B332, 100, 1994.

\refis{US} P. Bin\'etruy and P. Ramond, in preparation.

\refis{MR}S. Martin and P. Ramond, in preparation.

\refis{unification}
U.~Amaldi, W.~de Boer, and H.~Furstenau,
\pl B260, 447, 1991;
J.~Ellis, S.~Kelley and D.~Nanopoulos,
\pl 260B, 131, 1991;
P.~Langacker and M.~Luo,
\pr D44, 817, 1991.

\refis{gmrsy}M. Gell-Mann, P. Ramond, and R. Slansky in Sanibel
Talk,
CALT-68-709, Feb 1979, and in {\it Supergravity} (North Holland,
Amsterdam 1979). T. Yanagida, in {\it Proceedings of the Workshop
on Unified Theory and Baryon Number of the Universe}, KEK, Japan,
1979.

\refis{kounnas} J.P. Derendinger, S. Ferrara, C. Kounnas, and F. Zwirner, \np
B372, 145, 1992.

\refis{GKMR} B.~R.~Greene, K.~H.~Kirklin, P.~J.~Miron, and G.~G.~Ross,
\pl B180, 69, 1986; \np B278, 667, 1986; \np B292, 606, 1987.

\endreferences\endit\end